\documentclass[12pt,preprint]{aastex}

\def\aa#1#2#3#4#5{\bibitem[#1]{#2}#3, { A\&A}, {#4}, #5}

\def\apj#1#2#3#4#5{\bibitem[#1]{#2}#3, { Ap. J.}, {#4}, #5}

\def\pasp#1#2#3#4#5{\bibitem[#1]{#2}#3, { PASP}, {#4}, #5}

\def\science#1#2#3#4#5{\bibitem[#1]{#2}#3, Science, #4, #5}

\def\spie#1#2#3#4#5{\bibitem[#1]{#2}#3, Proc. SPIE, #4, #5}

\begin{document}

\title{Detecting Outer Planets in Edge-On Orbits:
Combining Radial Velocity and Astrometric Techniques}

\author{J. A.  Eisner \& S. R. Kulkarni}
\affil{Palomar Observatory 105-24, California Institute of
Technology, Pasadena, CA 91125}

\begin{abstract}
The astrometric and radial velocity techniques of extra-solar
planet detection attempt to detect the periodic reflex motion of
the parent star by extracting this periodic signal 
from a time-sampled set of observations.  
The extraction is generally accomplished
using periodogram analysis or the functionally equivalent 
technique of Least Squares fitting of sinusoids.  In this paper, 
we use a Frequentist approach
to examine the sensitivity of Least Squares technique when 
applied to a combination of radial velocity and astrometric 
observations.  We derive a semi-analytical expression for the sensitivity 
and show that the combined approach yields significantly better sensitivity
than either technique on its own. We discuss the ramifications
of this result to upcoming astrometric surveys with FAME, 
the Keck Interferometer, and SIM.

\end{abstract}

\section{Introduction}
Radial velocity (RV) surveys of nearby stars have been employed in the search
for extra-solar planets for nearly two decades 
(see Marcy, Cochran \& Mayor 2000). As these efforts continue
into the next decade, they will be supplemented by precision astrometric
searches, {e.g.,} by FAME, Keck Interferometer, and SIM \citep{HORNER+99,
VANBELLE+98,DU99}.
In previous papers, 
we examined the RV and astrometric techniques
in detail, paying particular attention to the regime where the
time-baseline of the observations is shorter than the orbital period
of the extra-solar companion (Eisner \& Kulkarni 2001a,b; 
hereafter EK2001a,b).  
This regime is interesting because  one expects giant
planets to form in the colder regions of the proto-planetary
nebula, and thus one expects such objects to possess periods of
many years to centuries \citep{BOSS95}.  In EK2001a,b
we demonstrated that one can achieve
a significant improvement in sensitivity (over current techniques)
if the orbital amplitude
{\em and phase} are included in the analysis.

Here, we examine the benefits of combining simultaneous
astrometric and RV observations.  Specifically,
we examine the sensitivity of a combined astrometric and RV detection
technique applied to an edge-on orbit, where the full RV signature and
one dimension of the astrometric signature can be observed.  

The
plan for the paper is straightforward. First, we simulate
large numbers of hypothetical data sets containing (1) noise only, and
(2) signal and noise, and determine the Frequentist Type I and II errors.
As in EK2001a,b we acknowledge that a Frequentist approach is not as
rigorous as a full Bayesian analysis.  However, this approach is simple
enough that it is amenable to deriving (semi-)analytical estimates of
the sensitivity -- a principal goal of the paper. We conclude by discussing
the parameter space opened up by combining FAME, Keck Interferometer, or SIM
astrometric surveys with ongoing precision RV studies.

\section{Basic Equations \label{sec:equations}}
We will assume edge-on circular orbits throughout this discussion.  
The astrometric signature of an edge-on circular orbit is given by
\begin{equation}
\theta(t) = {\cal A} \sin \left( \frac{2\pi t}{\tau} +\phi\right)
+ \lambda t + \mu,
\label{eq:astrom-orbit}
\end{equation}
where $\lambda$ and $\mu$ are the proper motion and parallax of the
planetary system, respectively, and
\begin{equation}
{\cal A} = \frac{M_p}{D}
\left(\frac{G \tau^2}{4\pi^2 M_{\ast}^2} \right) ^{\frac{1}{3}}. 
\label{eq:astrom-amplitude}
\end{equation}
Here, $D$ is the distance to the system, 
$M_{\ast}$ is the mass of the star, and $M_p$ is the mass of the
planet. We ignore the annual parallax.  However, annual parallax should
be included in modeling of planets with periods around one year.

The RV signature of this orbit is given by the derivative of the 
orbital position  along the line of sight:
\begin{equation}
v(t) = {\cal V}
\sin \left( \frac{2\pi t}{\tau} +\phi\right) + \gamma.
\label{eq:rv-orbit}
\end{equation}
Here $\gamma$ is the radial velocity of the planetary system, and
\begin{equation}
{\cal V} = M_p \left(\frac{2\pi G} {M_\ast^2\tau} \right)^{\frac{1}{3}}
= \frac{2\pi D}{\tau} \times {\cal A}.
\label{eq:rv-amplitude}
\end{equation} 
Thus, we can express the sensitivity (defined as the minimum-mass planet
that can be detected) of the RV and astrometric techniques in terms
of $\cal A$:
\begin{equation}
M_p = {D {\cal A}} \left(\frac{4\pi^2 M_{\ast}^2}{G \tau^2}\right)
^{\frac{1}{3}} .
\label{eq:sensitivity}
\end{equation}
However, it is more difficult to identify planets with long periods than 
Equation \ref{eq:sensitivity} might suggest.  In the so-called
``long-period regime'', defined as $T_0<<\tau$ where $T_0$ is the
duration of the survey, we observe
a fraction of the orbit.  As a result,
in this regime, the sensitivity is expected to
depend critically on the orbital phase.  
The reflex velocity is covariant with $\gamma$ and thus the RV
technique is most sensitive when 
$2\pi t/\tau + \phi = n\pi$ (EK2001a).  In contrast, the
astrometric signal of an edge-on orbit is covariant with $\lambda t$ and
$\mu$, and thus the astrometric technique is sensitive when 
$2\pi t/\tau + \phi = (n+1/2) \pi$ (EK2001b).  Thus the RV and astrometric
techniques achieve their maximal sensitivities for different
orbital phases, and we expect, on general grounds,
that combining the two techniques
should yield a substantial benefit in the long-period regime.

\section{Monte Carlo Analysis \label{sec:errors}}
The signal analysis for the astrometric and RV techniques consists
of fitting the observations to the models specified in
Equations \ref{eq:astrom-orbit} and \ref{eq:rv-orbit}.
As noted by several authors \citep[{e.g.}][EK2001a]{SCARGLE82,NA98} 
the most optimal fitting is obtained by using the
technique of Least Squares.  First, we convert the physical
model specified by Equations
\ref{eq:astrom-orbit} and \ref{eq:rv-orbit}
to equations linear in the unknowns:
\begin{equation}
\theta(t) = {\cal A}_c \cos(\omega t) + {\cal A}_s \sin(\omega t) + 
\lambda t + \mu,
\label{eq:lsq-model1}
\end{equation}
\begin{equation}
v(t) = {\cal V}_c \cos(\omega t) + {\cal V}_s \sin(\omega t) + \gamma.
\label{eq:lsq-model2}
\end{equation}
Here, ${\cal A}_c = {\cal A} \sin \phi$, ${\cal A}_s = {\cal A} \cos \phi$, 
${\cal V}_c = {\cal V} \sin \phi$, 
${\cal V}_s = {\cal V} \cos \phi$, and $\omega=2\pi/\tau$.  
In EK2001a,b we discuss
the importance of the $\gamma$, $\lambda$ and $\mu$ terms.
These three variables are not directly relevant in detecting or characterizing
a companion planet but they are unknown and in the long-period
regime are covariant with some of the orbital parameters
\citep{BS82}. Thus
the three variables must be solved for in order to correctly model the
observations. 

Using Equations \ref{eq:lsq-model1} and \ref{eq:lsq-model2} as our
physical model, we perform the following analysis.  First, we simulate
a large number of data-sets containing only Gaussian nose ({i.e.}
no signal).  For each of these data sets, we perform a Least Squares
fit to three models: a model using only astrometric measurements
(Equation \ref{eq:lsq-model1}), a model using only RV measurements
(Equation \ref{eq:lsq-model2}), 
and a model that utilizes both astrometric and
RV measurements. In each case, for each simulated data set we fit for
amplitude and phase. We note here that for the RV+astrometry model,
since the two measurements have different variances we minimize the
$\chi^2$ (where $\chi$ is the difference between the model
and the rms-weighted measurements).

Specifically, we simulate $N = 1000$ data-sets, sampled at one month
intervals for $T_0 = 10$ years (with no loss of generality, we take the
time interval to go from $-T_0/2$ to $T_0/2$), and we explore
periods from 5 to 100 years.   We assume that the measurement noise in
both the RV and astrometric surveys is characterized by Gaussian noise
with rms of $\sigma_{\rm RV}$ and $\sigma_{\rm ast}$ respectively. The
best achieved $\sigma_{\rm RV} = 3$ m s$^{-1}$ \citep{BUTLER+96}.
The anticipated astrometric precision of FAME is between 50 and 100 $\mu$as 
\citep{HORNER+99}, that of the Keck Interferometer (narrow angle) 
between $30$ and 50 $\mu$as \citep{VANBELLE+98}, and that of SIM
between 1 and 10 $\mu$as \citep{DU99}. 
We note that $\sigma_{\rm ast} = 100$ $\mu$as yields
approximately equivalent sensitivity to RV technique with 3 m s$^{-1}$
rms for a planet orbiting a star located at distance $D=10$ pc with
$\tau \sim 2\: T_0 \sim 20$ years (Equation \ref{eq:rv-amplitude}).

Next, for each of three models, we determine the ellipse (in $\cal A$--$\phi$
space) within which 99\% of the fitted amplitudes and phases lie.
This ellipse, denoted by $\epsilon_1$, describes the ``Type I'' errors
of the detection technique.  Thus the inferred $\cal A$ and $\phi$ have a
1\% chance of being outside the $\epsilon_1$ ellipse (in the absence of
a signal).

\begin{figure}
\plotone{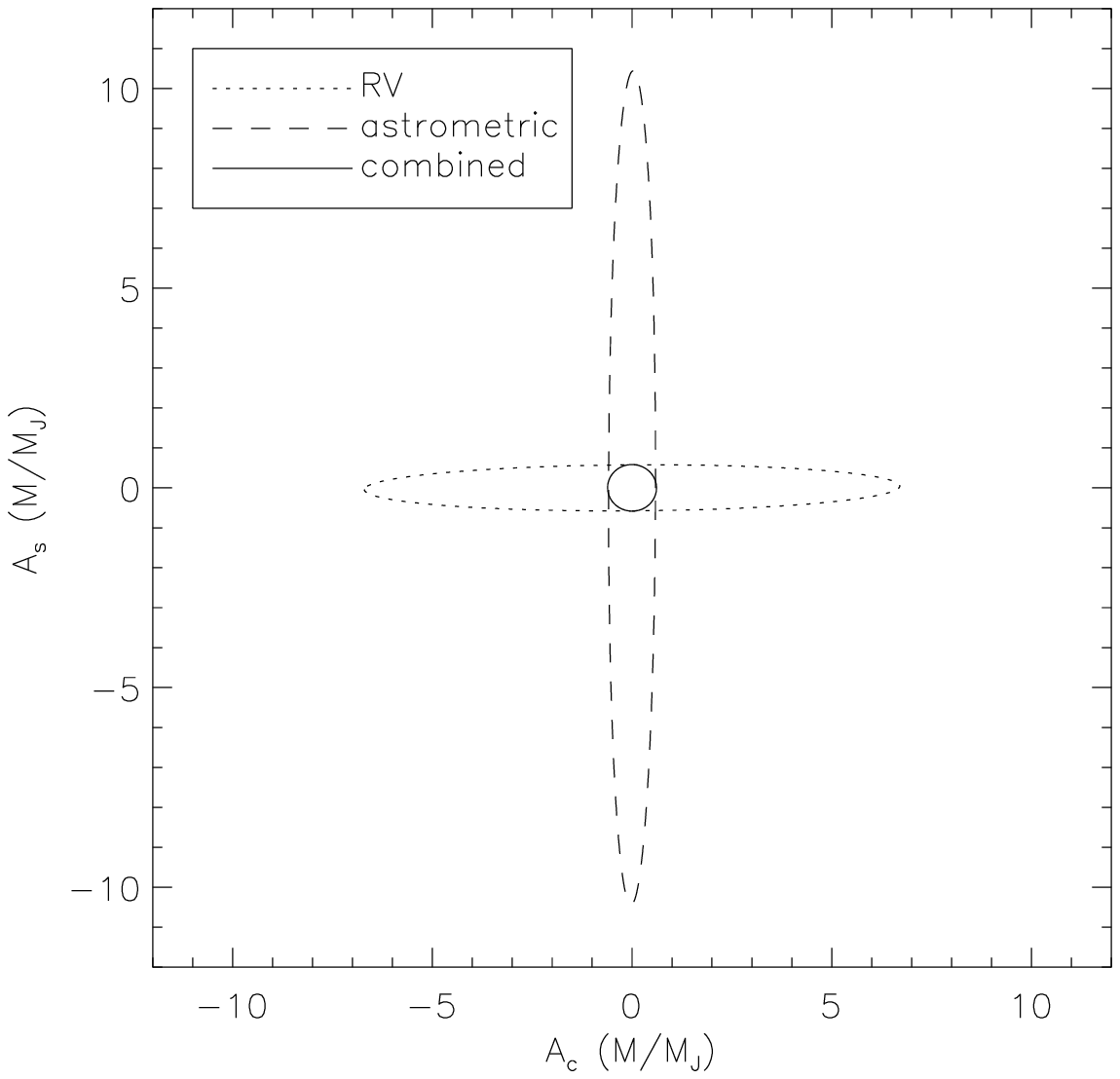}
\caption{A plot of the 1\% error ellipses for the astrometric and
RV techniques for a 90 year orbit ($\tau \sim 9 \: T_0$).  The
error ellipse for combined RV+astrometry technique is also shown.
These are the ellipses, $\epsilon_1$, for which 99\% of Least
Squares fits to simulated Gaussian noise produce fitted
amplitudes and phases that lie within $\epsilon_1$ (\S \ref{sec:errors}).
We have scaled the ellipses to units of companion mass
via Equation \ref{eq:sensitivity}.
\label{fig:ellipses}}
\end{figure}

As discussed earlier (\S\ref{sec:equations}) we expect RV and astrometric
models to show orthogonal sensitivity. Indeed, as can be seen from Figure
\ref{fig:ellipses}, the $\epsilon_{\rm 1, RV}$ and $\epsilon_{\rm 1,
ast}$ are $90^{\circ}$ out of phase in the long period regime.  On a basic
level, we can understand the benefit of combining RV and astrometric
observations by noting that the intersection of $\epsilon_{\rm 1,
rv}$ and $\epsilon_{\rm 1, ast}$ is much smaller than either of the
individual ellipses, and thus it is easier to detect signals over the
level of the noise.  In fact, this is verified by the combined analysis:
the ellipse for the combined analysis, $\epsilon_{1, C}$, lies entirely
within {\em both} $\epsilon_{\rm 1, RV}$ and $\epsilon_{\rm 1, ast}$
(Figure \ref{fig:ellipses}).

\begin{figure}
\plotone{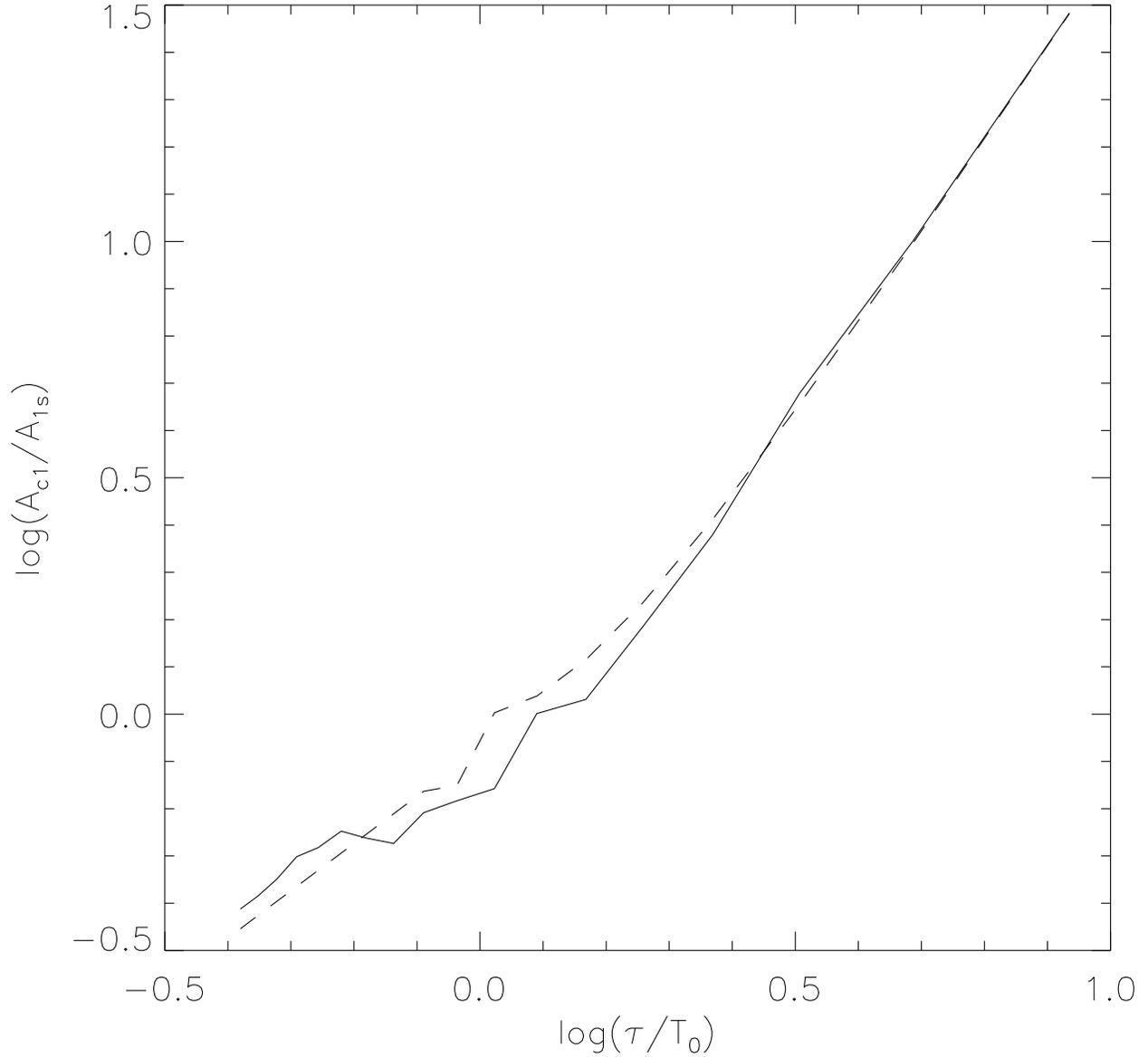}
\caption{A plot of $A_{c1}$ versus orbital period (solid line).  
The analytic expression given by Equation \ref{eq:ac} is also
plotted (dashed line).  $A_{c1}$
is the value of $\vert A_c \vert$ that is exceeded in 1\% of
least-squares fits to Gaussian noise, and 
describes the radius of the Type I error ellipse (actually a circle
in this case).
\label{fig:ac}}
\end{figure}

Analytic expressions for the Type I errors for RV or astrometric
techniques are given in EK2001a and EK2001b, respectively.  Given
these expressions, it is not difficult to infer an analytic
expression for the Type I errors in the case of combined
astrometric+RV technique.
As noted earlier 
for $\sigma_{\rm RV}=3$ m s$^{-1}$ and $\sigma_{\rm ast}=100$ $\mu$as, 
the semi-minor axes for $\epsilon_{\rm 1,RV}$ and $\epsilon_{\rm 1,ast}$
are approximately
equal ($D=10$ pc), and thus $\epsilon_1$ for the combined analysis will be
a circle whose radius is given by
\begin{equation}
A_{c1} = \cases{2^{-1/2}\: {\rm min}(A_{1s},\frac{\tau}{2\pi D}v_{1s}) 
& for $\tau < T_0$ \cr
\frac{2 A_{1s}}{1-\cos(\pi T_0/\tau)} & for $\tau > T_0$}.
\label{eq:ac}
\end{equation}
Here, $A_{1s} = 3.69 \sigma_{\rm ast} n_0^{-1/2}$, $v_{1s}= 3.69
\sigma_{\rm rv} n_0^{-1/2}$, and the factor of $2^{-1/2}$ reflects the
fact that in the short-period regime, there are essentially twice as
many measurements.  As illustrated in Figure \ref{fig:ac}, this analytic
function provides an excellent fit to the data.

Next, we evaluate ``Type II'' errors for the three models.  Type II
errors describe the probability of failing to detect a genuine signal
due to contamination by noise.  To understand the type II statistics,
we simulate a large number of data sets consisting of a simulated signal
and noise (see EK2001a,b for further details).
The signal is a sinusoidal wave with an amplitude ${\cal A}_0$ (astrometry),
and the corresponding velocity amplitude is $2\pi D {\cal A}_0/\tau $ (we
set $D=10$ pc); the phase, $\phi$ is randomly chosen from the interval
$[0,2\pi]$ (uniform distribution). For each model, 
we increment the amplitude[s]
until 99\% of the fitted orbital parameters lie outside of the appropriate
$\epsilon_{\rm 1}$ ellipse; this amplitude is denoted by $A_{99}$
(for each method).

\section{Results and Discussion \label{sec:results}}

The benefit of combining RV and astrometric analysis
accrues mainly from the fact that the error ellipses for the two techniques
in $A-\phi$ parameter space are perpendicular to each other
(Figure \ref{fig:ellipses}). 
RV+astrometry analysis will be most useful in cases where
the error ellipses for the two techniques are roughly the same size
(otherwise, one error ellipse might lie entirely within the other,
and no additional benefit would arise from combining the two
techniques).  As mentioned above, the current precision of RV
techniques is $\sim 3$ m s$^{-1}$ \citep{BUTLER+96}, which means
that for a 10 year survey,
we must use astrometric measurements with $\sim 100$ $\mu$as
precision (for a system at $D=10$ pc) to reap the maximal benefit
from RV+astrometry technique.  This is approximately the sensitivity
that will be obtained by future instruments like Keck Interferometer
and FAME \citep{VANBELLE+98,HORNER+99}.  

\begin{figure}
\plotone{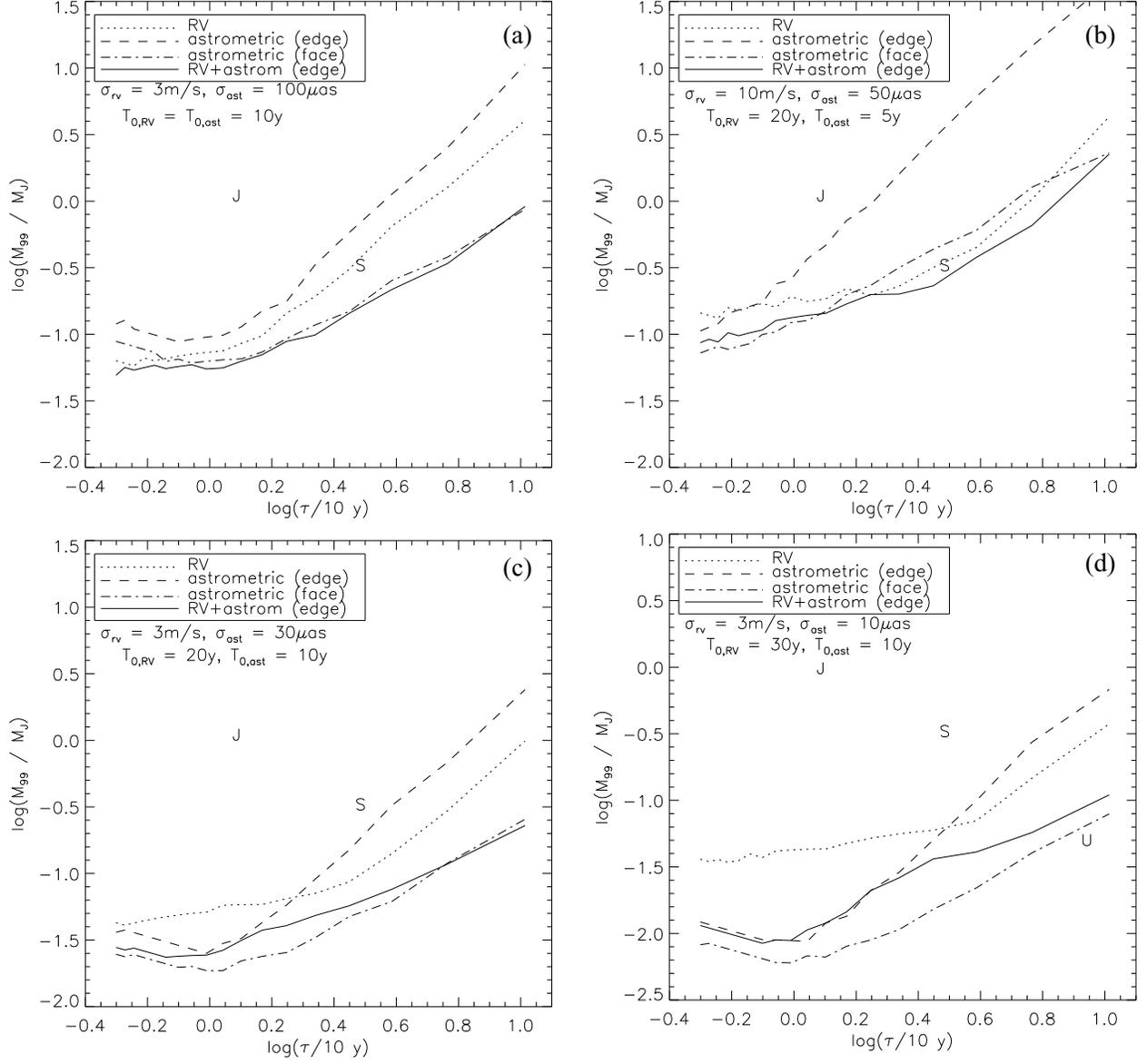}
\caption{Plots of $\log(M_{99})$, in units of Jupiter masses, versus
$\log (\tau / T_0)$.  $M_{99}$ is
$A_{99}$ expressed in units of companion mass, assuming
$M_{\ast} =M_{\odot}$ (Equation \ref{eq:sensitivity}).
The positions of Jupiter, Saturn and Uranus in this parameter space 
are also indicated. In (a), both the RV and astrometric surveys
have a duration of  $T_0=10$ years.  (b) Shows the simulated sensitivity
for FAME+RV, (c) shows Keck Interferometer+RV, and (d) shows SIM+RV.
\label{fig:m99}}
\end{figure}

As illustrated by Figure \ref{fig:m99}a, 
RV+astrometry analysis
(with comparable RV and astrometric measurement accuracies)
applied to edge-on orbits attains approximately the same sensitivity
as an astrometric analysis applied to face-on orbits.  This similarity
stems from the fact that in both cases, the highly elliptical 1-D  $\epsilon_1$
is circularized through the addition of a second dimension.  Another way of 
thinking about this is that no matter what part of the orbit, when we observe
both dimensions we can always see the full orbital curvature.  Thus,  
combining astrometric and RV techniques in a large
survey ensures good sensitivity for all orbital inclination angles.

It is also worth noting that RV+astrometry yields valuable gains in
the short-period regime ($\tau < T_0$).  When $\sigma_{\rm RV}$ is
comparable to $\sigma_{\rm ast}$, the sensitivity of RV+astrometry
is better by $2^{-1/2}$ over RV or astrometry alone.  
Furthermore, noting that 
the sensitivity of astrometry to face-on orbits is $2^{-1/2}\sigma_{\rm ast}$
(EK2001b), we see that the short-period sensitivity of RV+astrometry is
approximately independent of orbital inclination.

We have also examined the sensitivity when astrometric data
is combined with RV data of a longer time-baseline,
specifically for several upcoming
missions (recall that RV surveys
will have been underway for 15--20 years by the time astrometric surveys
commence).  We investigate FAME
($T_{0, \rm RV} = 20$ years, $T_{0, \rm ast} = 5$ years), Keck Interferometer
($T_{0, \rm RV} = 20$ years, $T_{0, \rm ast} =10$ years), and SIM
($T_{0, \rm RV} = 30$ years, $T_{0, \rm ast} =10$ years).  
We find that in the cases of FAME and Keck Interferometer, the addition
of longer time-baseline RV measurements has a significant impact
(Figures \ref{fig:m99}b--\ref{fig:m99}c).  In fact,
RV+astrometry analysis can easily detect Saturns when astrometric analysis
alone doesn't come close (Figure \ref{fig:m99}b).  
In the case of SIM, the main
benefit of RV+astrometry is that one can achieve optimal
sensitivity over a wider range of inclination angles (Figure \ref{fig:m99}d).

\subsection{Shorter Surveys}
We have also examined the sensitivities of the RV, astrometric,
and combined techniques
applied to shorter duration surveys, in order to compare the various
sensitivities for short-period companions.  
Specifically, we investigate the prospect of finding companions
around M dwarfs with a 2-year survey combining
RV measurements with astrometric measurements from AO systems on large
telescopes like Keck or the Palomar $200''$.  Previous authors
have successfully searched for companions around M-dwarfs using RV measurements
and astrometric measurements on AO systems \citep[{e.g.,}][]{DELFOSSE+99}
although they have not used the combined RV+astrometry analysis described here
({i.e.,} they analysed the RV data and the astrometric data separately).

\begin{figure}
\plottwo{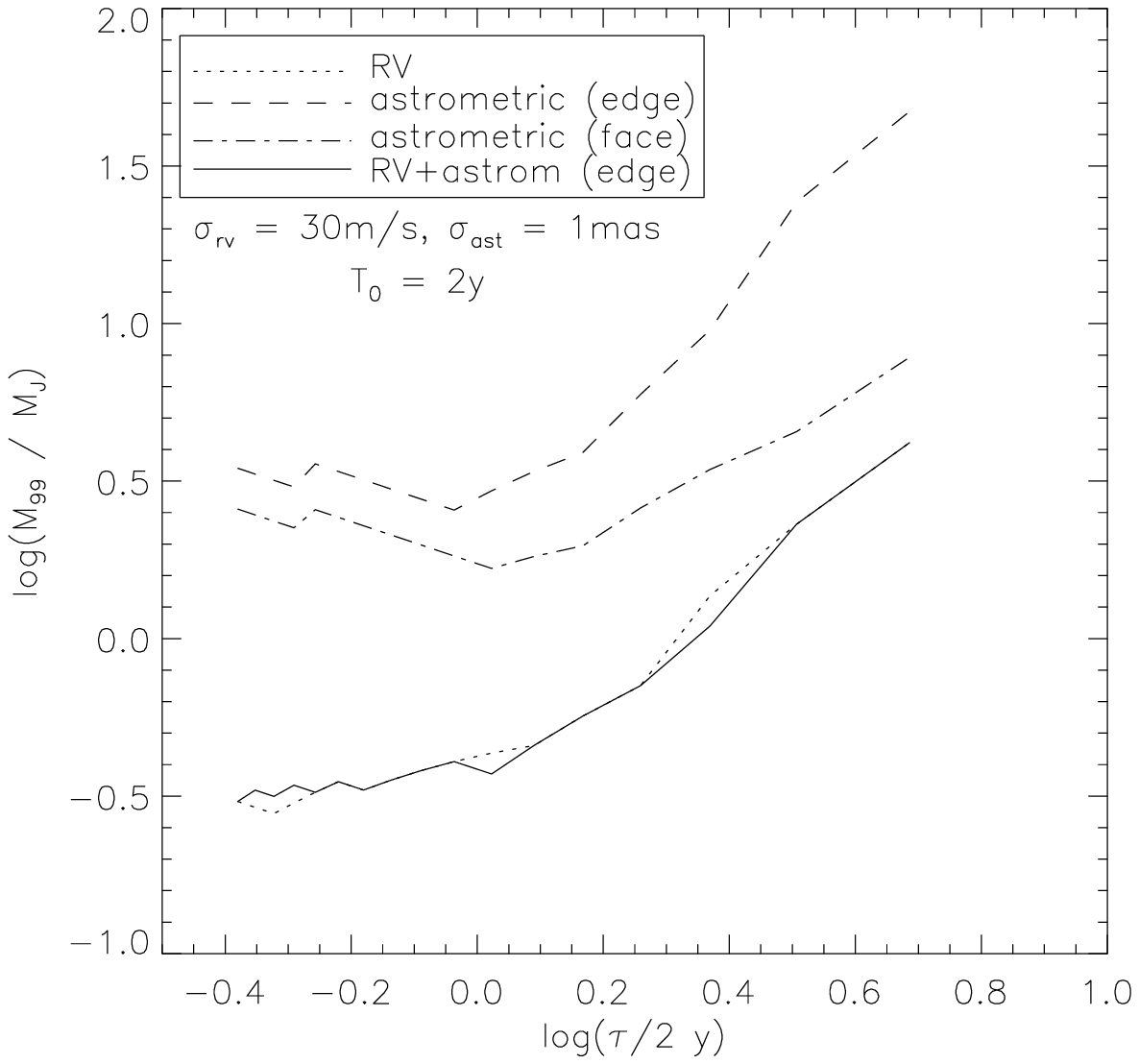}{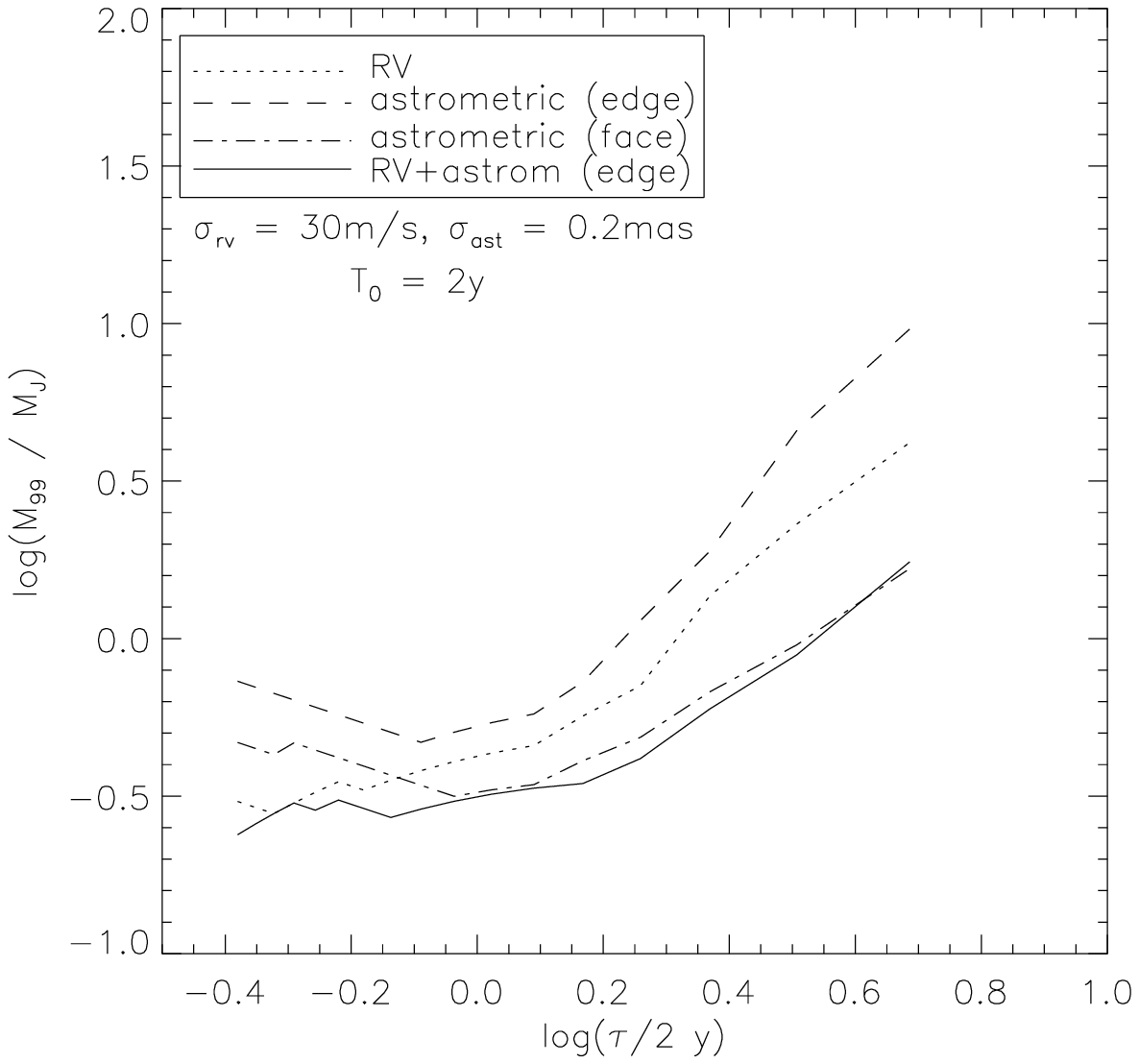}
\caption{Plots of $\log(M_{99})$, in units of Jupiter masses, versus
$\log (\tau / T_0)$.  $M_{99}$ is
$A_{99}$ expressed in units of companion mass,
assuming $M_{\ast} = 0.25 M_{\odot}$ (Equation \ref{eq:sensitivity}).
\label{fig:bd}}
\end{figure}

As illustrated by Figure \ref{fig:bd}, the RV technique gains 
significantly over astrometry for short periods
(because ${\cal V} \propto {\cal A}/\tau$; Equation \ref{eq:rv-amplitude}).
If astrometry is to contribute meaningfully then $\sigma_{\rm ast}
\lesssim 1$ mas.  There is some expectation that such a precision
can be obtained for binary stars \citep{DEKANY+94}.  If so,
combined RV+Astrometry surveys of nearby M dwarfs can measure masses
of Jupiter and Saturn-mass companions.

\end{document}